\documentclass[twoside]{ilcws08}
\usepackage[latin1]{inputenc}
\usepackage[dvipdfm]{graphicx,epsfig,color}
\usepackage{wrapfig,rotating}
\usepackage{amssymb,amsmath,array}
\begin{document}
%%%%   Paper title goes here  %%%%%%%%%%%%%%
\title{SUSY-P5: Chargino / Neutralino Analysis in the Fully Hadronic Final State}
%%
%***********************************************************************
% AUTHORS INFORMATION AREA
%***********************************************************************
\author{
  Daniela K\"afer$^1$, 
  Jenny List$^1$, and 
  Taikan Suehara$^2$
  % short acknowledgment:
  \thanks{Authors K\"afer and List acknowledge the support by DFG Li 1560/1-1, 
    and author Suehara acknowledges the support by KAKENHI 18GS0202.}
  %% ------------------------------------------------------------
  % DO NOT MODIFY THE FOLLOWING '\vspace' ARGUMENT
  \vspace{.3cm}\\
  %% 
  %% ---------------------------------------------------------------------
  1- Deutsches Elektronen Synchrotron (DESY) - Hamburg, Germany \\[0.4mm]
  2- International Center for Elementary Particle Physics (ICEPP),
  Univ. of Tokyo - Tokyo, Japan \\[-1.4mm]
  %% ---------------------------------------------------------------------
}
%% ***********************************************************************
% END OF AUTHORS INFORMATION AREA
%***********************************************************************

%%%%%%%%%%%%%%%%%%%%%%%%%%%%%%%%%%%%%%%%%%%%%%%%%%%%%%%%%%%
%%   own commands  
%%%%%%%%%%%%%%%%%%%%%%%%%%%%%%%%%%%%%%%%%%%%%%%%%%%%%%%%%%%
\newcommand{\neutone}  {\ensuremath{ \widetilde{\chi}_1^0     }}
\newcommand{\neuttwo}  {\ensuremath{ \widetilde{\chi}_2^0     }}
\newcommand{\charone}  {\ensuremath{ \widetilde{\chi}_1^{\pm} }}
\newcommand{\charonemp}{\ensuremath{ \widetilde{\chi}_1^{\mp} }}
%---------------------------------------------------------------
\newcommand{\Ra}      {\ensuremath{ \Rightarrow     }}
\newcommand{\longra}  {\ensuremath{ \longrightarrow }}
\newcommand{\Longra}  {\ensuremath{ \Longrightarrow }}
%-----------------------------------------------------
\newcommand{\la}      {\ensuremath{ \leftarrow      }}
\newcommand{\La}      {\ensuremath{ \Leftarrow      }}
\newcommand{\longla}  {\ensuremath{ \longleftarrow  }}
\newcommand{\Longla}  {\ensuremath{ \Longleftarrow  }}
%-----------------------------------------------------
\newcommand{\lra}     {\ensuremath{ \leftrightarrow }}
\newcommand{\Lra}     {\ensuremath{ \Leftrightarrow }}
\newcommand{\longlra} {\ensuremath{ \longleftrightarrow }}
\newcommand{\Longlra} {\ensuremath{ \Longleftrightarrow }}

%%-------------------------------------------------------------------------------------------
\definecolor{mgrey}      {rgb}{0.45, 0.45, 0.45}  \newcommand{\mgrey}      {\color{mgrey}}
\definecolor{dgrey}      {rgb}{0.28, 0.28, 0.28}  \newcommand{\dgrey}      {\color{dgrey}}
%%-------------------------------------------------------------------------------------------
\definecolor{dred}       {rgb}{0.76, 0.00, 0.00}  \newcommand{\dred}       {\color{dred}}
\definecolor{dgreen}     {rgb}{0.05, 0.50, 0.10}  \newcommand{\dgreen}     {\color{dgreen}}
\definecolor{mgreen}     {rgb}{0.10, 0.90, 0.10}  \newcommand{\mgreen}     {\color{mgreen}}
%%-------------------------------------------------------------------------------------------
\definecolor{mblue}      {rgb}{0.20, 0.36, 0.72}  \newcommand{\mblue}      {\color{mblue}}
\definecolor{dblue}      {rgb}{0.02, 0.05, 0.54}  \newcommand{\dblue}      {\color{dblue}}
%%-------------------------------------------------------------------------------------------
%%%%%%%%%%%%%%%%%%%%%%%%%%%%%%%%%%%%%%%%%%%%%%%%%%%%%%%%%%%

\maketitle

\begin{abstract}
  %%----------------------------------------------------------
  The fully hadronic final states of two signal processes from an mSUGRA 
  inspired scenario (SUSY-P5) %~\cite{bib:SUSY-P5} 
  are studied within a full simulation of the LDC$^{\prime}$ detector model. 
  These are chargino pair and neutralino pair production, 
  %%   e+e- -> C1-C1+ -> qq'N1 qq'N1 and  e+e- -> N2N2  -> qqN1qqN1. 
  i.e. $e^+e^-\to \charone\charonemp \to q\bar{q}^{\prime}\neutone \; q\bar{q}^{\prime}\neutone$ 
  and  $e^+e^-\to \neuttwo\neuttwo   \to q\bar{q}\neutone         \; q\bar{q}\neutone$.
  Both processes have to be separated sufficiently from all background to measure 
  the respective production cross sections and extract the masses of the involved 
  bosinos, $m(\charone)$, $m(\neuttwo)$ and $m_{LSP}=m(\neutone)$. This is achieved 
  by fitting the energy spectra of the reconstructed gauge bosons while taking 
  into account the finite width of the boson mass. %  and beam energy spread.  
  From simulation data corresponding to 500~fb$^{-1}$ of luminosity, % (signal processes), 
  a mass resolution of about 0.5 GeV seems to be achievable.
  %%----------------------------------------------------------
\end{abstract}

%%===========================================================================
%% page 1
%%========
\section{Introduction}
Due to the clean environment and precisely known inital state at the ILC, it will 
be possible to measure very small signal cross sections and extract the masses 
of SUSY particles even from fully hadronic final states. As an example, from the 
SUSY-P5 scenario~\cite{bib:SUSY-P5}, pair production of charginos ($\charone\charonemp$) 
and neutralinos ($\neuttwo\neuttwo$) and their subsequent hadronic decays 
($q\bar{q}\neutone\; q\bar{q}\neutone$) are studied. 
To successfully extract the masses of intermediate SUSY particles from a fully 
hadronic final state, it is mandatory to reduce the background to a minimum. 
Especially difficult is the  mutual separation of both signal processes. 
Four jets need to be assigned to two gauge bosons, either two $W$ or two $Z$ 
bosons, depending on whether they originated from a decaying chargino or neutralino. 
In addition, a considerable amount of missing energy and momentum carried away 
by two of the lightest supersymmetric particles (LSPs) impede a complete 
reconstruction of the final state.

All Monte Carlo samples are part of the DESY mass production~\cite{bib:DESYsamples}. 
The events are generated with \texttt{Whizard}~\cite{bib:Whizard}, run through 
the full simulation of the \texttt{LDCPrime\_02Sc} detector model and finally 
reconstructed with \texttt{MarlinReco}~\cite{bib:MarlinReco} using Pandora 
Particle Flow (PFlow)~\cite{bib:PFlow}.

\section{Selection strategies and kinematic fits}
\begin{figure}[!h]
  \setlength{\unitlength}{1.0cm}
  \begin{picture}(10.0, 4.43)
    % ---------------------
    \put( 0.00,-0.2)  {\epsfig{file=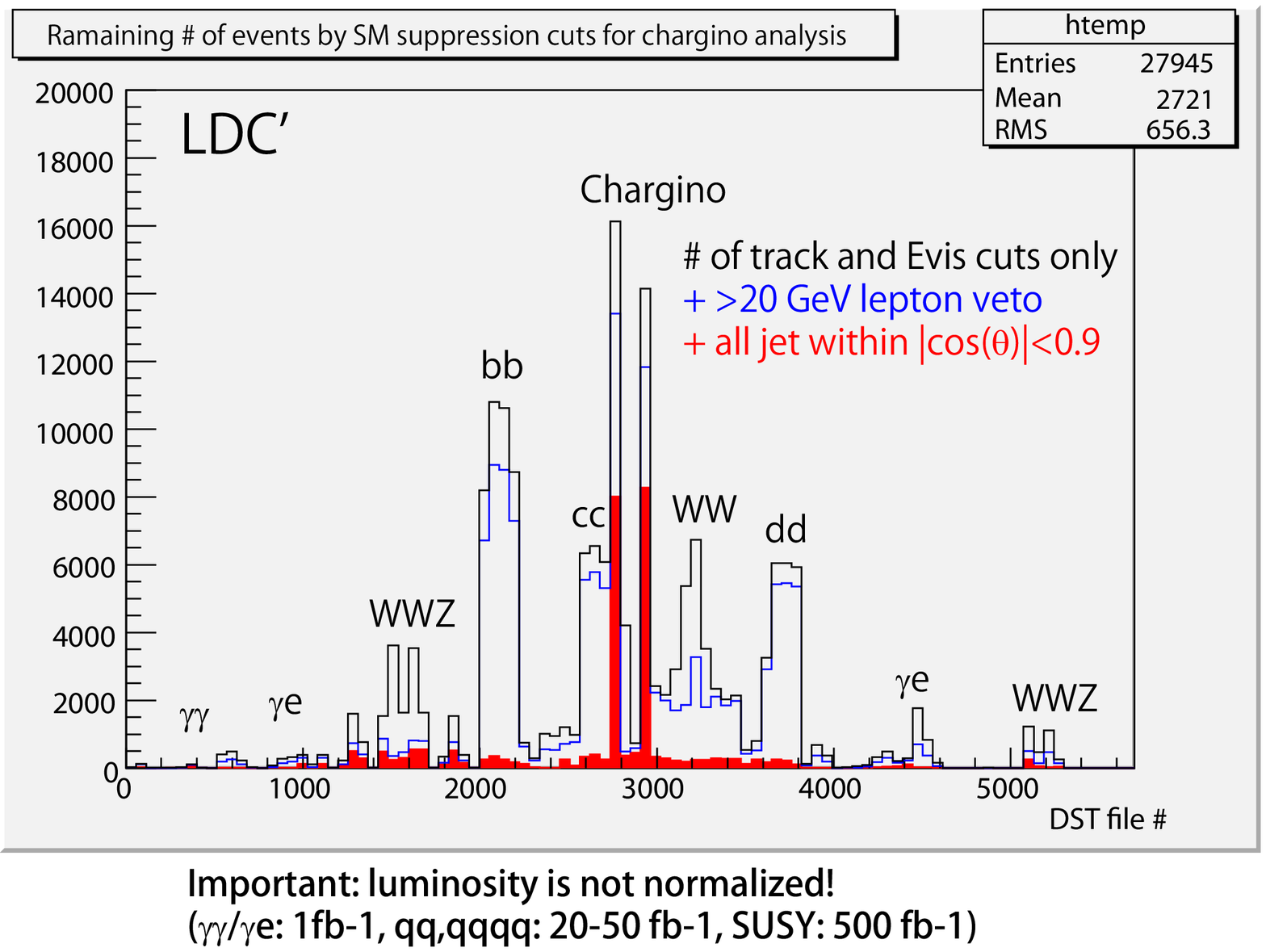, bb= 0 45 567 426, clip= , width=0.50\linewidth}}
    \put( 7.05,-0.2)  {\epsfig{file=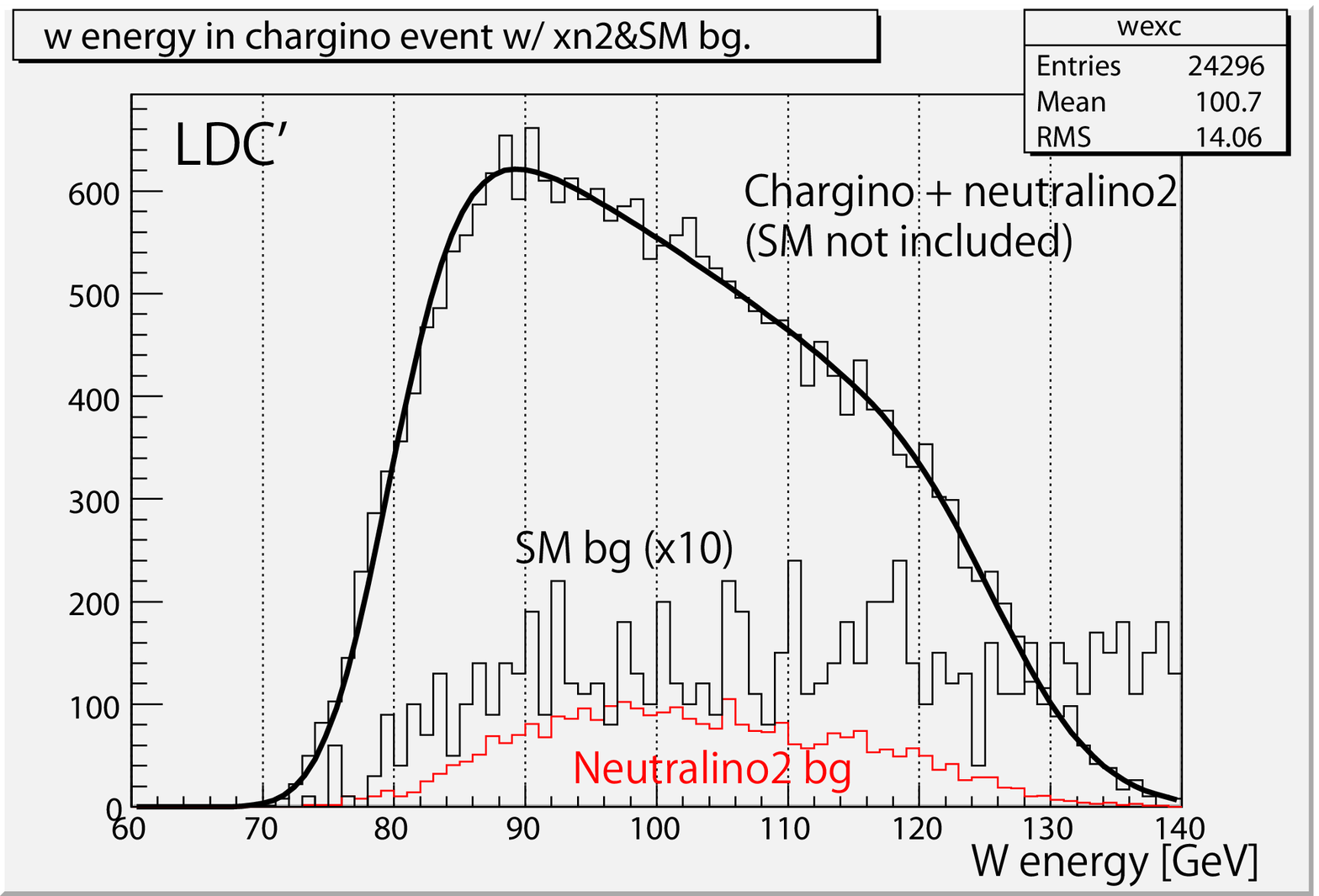, bb= 0  0 567 385, clip= , width=0.50\linewidth}}
    % ---------------------
    \put( 0.02,-0.1)  {(a)}
    \put( 7.07,-0.1)  {(b)}
    % ---------------------
    \put( 8.20, 2.5)  {\scriptsize\dblue $m(\charone)=219.7 \pm 1.54$~GeV}
    \put( 8.20, 2.1)  {\scriptsize\dblue $m(\neutone\,)\,=121.2 \pm 0.59$~GeV\qquad (LSP)}
    % C1 mass: 219.7 +- 1.54 GeV
    % N1 mass: 121.2 +- 0.59 GeV   (LSP)
  \end{picture}
  \caption{(a) SM background contributions. Events left after requiring: 
    black: $N_{tracks}>20$ and $150\;$GeV$\,<E_{vis}<\,$300$\;$GeV; 
    blue: lepton veto with $E_{lepton}<20$~GeV;
    red: $E_{jet}>5$~GeV and $|\cos(\theta_{jet})|<0.9$ for all jets.$\;$ 
    (b) Empirical fit to the energy spectrum of the $W$ boson consisting 
    of a $3^{rd}$-order polynomial convoluted with a Gaussian (see text).} 
  \label{fig:SMback-fitWene-ts} 
  % (a) smcuts1.png
  % x-axis: DST file number,   y-axis: passed number of events
  % for Evis+ntrack cut, +lepton veto cut, + angular cut, 
  % see: /scratch/data/kaefer/tex/talks/DESY/figures/taikan-lcws08/081112-ildopt-suehara.pdf
  % ----------------------------------------------------------
  % (b) xcfit-wosmbg.png 
  %     -> Chargino massfit of LDC'
  % 3rd polynomial (4 parameters) (center) / 0 (edge), 
  % convoluted with a Gaussian (2 par.  sigma: linear function of energy)
  % edge position:  2 par.
  % -->  total number of parameters: 8
  % Fitted values:
  % C1 mass: 219.7 +- 1.54 GeV
  % N1 mass: 121.2 +- 0.59 GeV   (LSP)
  % ----------------------------------------------------------
  % SM background: x10, because lumi is only 50 fb-1 for most backgrounds
  % Is not included in the fitted distribution.
  % (Unable to fit with SM background currently)
  % N2 background is included in the fitted distribution.
  % ----------------------------------------------------------
\end{figure}

This study concentrates on the reduction of SM background using a few very basic 
requirements and a kinematic fit to improve separation of the chargino from the 
neutralino signal. 
A rough background scan as depicted in Fig.~\ref{fig:SMback-fitWene-ts}~(a) 
shows $WW$ and $WWZ$ production to be the dominant source of SM background. 
Especially, 6-fermion final states from $WWZ$ production, with an invisible 
$Z$-decay, i.e. $q\bar{q^{\prime}}q\bar{q}^{\prime}\nu\bar{\nu}$, constitute 
a nearly irreducible background component. However, the processes shown in 
Fig.~\ref{fig:SMback-fitWene-ts} are not normalised to the same luminosity. 
While the SUSY signal processes correspond to 500~fb$^{-1}$, background samples were only 
%%===========================================================================
%% page 2
%%========
available for $\approx\,$20-50~fb$^{-1}$ for QCD multijet production 
($qq$, $qqqq$ final states) and 1~fb$^{-1}$ for $\gamma\gamma$/$e\gamma$ processes. 
Background from other SUSY processes has also not yet been considered, but it 
is expected to not render the preliminary results of this study obsolete. 

%------------------------------------
%------------------------------------

Some very basic requirements reduce leptonic events and events from multijet 
QCD production (see caption of Fig.~\ref{fig:SMback-fitWene-ts}), while the 
very similar neutralino signal is hardly affected. 
%%%%%%%%%%%%%%%%%%%%%%%%%%%%%%%%%%%%%%%%%%%
% The cut is based on two di-jet masses, comparison between chi2 value 
% of W-pair hypothesis and chi2 value of Z-pair hypothesis.
% 
% chi2(W,jetpair1)+chi2(W,jetpair2)<2 and
% chi2(Z,jetpair1)+chi2(Z,jetpair2)>4
% events are treated as chargino candidates.
% 
% chi2(W,jetpair) is defined as (mW - invmass_jetpair)^2 / (5 GeV)^2
% chi2(Z,jetpair) is defined as (mZ - invmass_jetpair)^2 / (5 GeV)^2
% 
% Jetpairs for W is selected (from 4 jets) so that sum of chi2 gives minimum, and
% Jetpairs for Z is selected (from 4 jets) so that sum of chi2 gives minimum.
% So, Jetpairs for W and Jetpairs for Z are sometimes different combination of jets.
%%%%%%%%%%%%%%%%%%%%%%%%%%%%%%%%%%%%%%%%%%%
To separate the chargino and neutralino signals from each other, two hypotheses 
are compared for the invariant di-jet masses, which can either form a $W$ or a 
$Z$ boson. The resulting $\chi^2$-values for each hypothesis are plotted against 
each other, so that the final selection criterium for chargino events results in: \\[-5mm]
\begin{align*}
  \chi^2(W,\, j_1j_2)+\chi^2(W,\, j_3j_4) &< 2 \hspace*{5mm}\mbox{with}\hspace*{3mm} &\chi^2(W,\, j_kj_l) = &\left(m_W - m(j_k,j_l)\right)^2/ (5\;\mbox{GeV})^2\hspace*{4mm}\\
  \chi^2(Z,\, j_1j_2)+\chi^2(Z,\, j_3j_4) &> 4 \hspace*{5mm}\mbox{with}\hspace*{3mm} &\chi^2(Z,\, j_kj_l) = &\left(m_Z - m(j_k,j_l)\right)^2/ (5\;\mbox{GeV})^2\,. 
\end{align*}
A more detailed description of the entire procedure is given in~\cite{bib:Taikan-Cambridge}. 
%%%%%%%%%%%%%%%%%%%%%%%%%%%%%%%%%%%%%%%%%%%
After the above selection of chargino-type events, the energy spectrum of 
di-jet masses in Fig.~\ref{fig:SMback-fitWene-ts}~(b) largely resembles 
that of the $W$ boson\footnote{SM background is multiplied by 10, 
  since so far the luminosity is only about 50 fb$^{-1}$ for most 
  background processes. It is also not included in the fit, while 
  the neutralino pair background is taken into account.}. 
The spectrum is fitted with an empirical function consisting of a $3^{rd}$-order 
polynomial (4+2 par. to describe the edge positions) convoluted with 
a Gaussian (2 par.) yielding the following values for the bosino masses: \\[-4mm]
\begin{align*}
  m(\charone) =\; & 219.7 \pm 1.54\;\,\mbox{GeV} \quad \lra \quad  216.5\;\,\mbox{GeV} \hspace*{6mm} \mbox{(nominal chargino mass)}       \\
  m(\neutone) =\; & 121.2 \pm 0.59\;\,\mbox{GeV} \quad \lra \quad  115.7\;\,\mbox{GeV} \hspace*{6mm} \mbox{(nominal neutralino/LSP mass)}.
\end{align*}
The central fit values differ rather much from the nominal bosino masses in the 
SUSY-P5 scenario. However, for an analysis of real data, such a shift can be 
compensated by tuning the MC distribution such that the experimental data is 
well reproduced.
%---------------------------------

\begin{figure}[!h]
  \setlength{\unitlength}{1.0cm}
  \begin{picture}(10.0, 3.9)
    \put(-0.10,-0.3)  {\epsfig{file=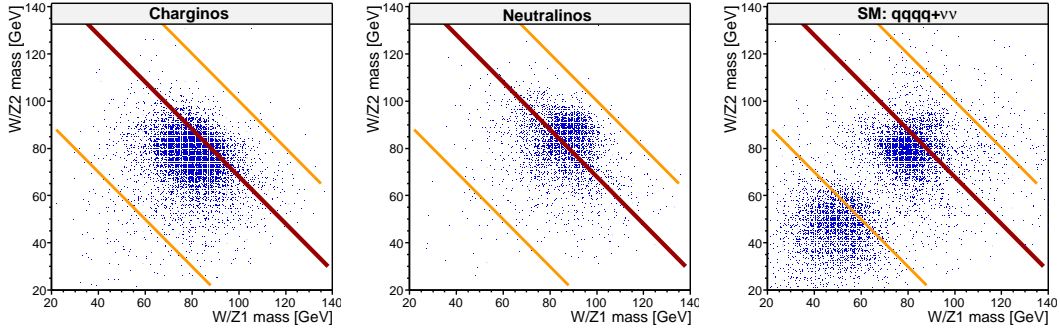, clip= , width=1.02\linewidth}}
  \end{picture}
  \caption{Selection of chargino-type events: all events between the middle (red) 
    and the lower (orange) lines are assumed to originate from chargino decays, 
    while events between the middle and upper lines are taken to be neutralino decays.}
  \label{fig:2dsel}
\end{figure}

A different approach postpones the reduction of events from SM background in 
favour of a slightly better separation of chargino and neutralino signals with 
the help of a 2-dim. cut in the ($V1,\,V2$) mass plane, with $V$ being a $W$ 
or $Z$ boson (see Fig.~\ref{fig:2dsel}). 
The cut is tuned, but not yet optimised, for a high efficiency and purity of 
the chargino pair production ($\varepsilon\times\pi$).

A simple kinematic 1C-fit, requiring equal di-jet masses (EMC), is applied to 
the three di-jet combinations of all events that can be clustered into 4 jets. 
The fit is meant to help find the correct jet-pairings, improve the mass and 
energy resolutions for the gauge bosons and thus, ultimately, the starting point 
for a fit to the $W$ ($Z$) boson energy spectra. 
The di-jet masses populating the ($V1,\,V2$) mass plane in Fig.~\ref{fig:2dsel} 
correspond to the jet-pairings with the best fit probability and are assumed to 
be the correct ones. 
%%===========================================================================
%% page 3
%%========
%------------------------------------

\section{Physics-driven fit to the energy spectrum of the {\boldmath$W$} boson}
Having selected the events between the lower two lines in Fig.~\ref{fig:2dsel} as 
chargino events, the energy spectrum of the corresponding gauge boson is then fitted 
with the masses of the SUSY particles as free parameters. 
However, this time, not an empirical fit function is used but a physics-driven function.  
Firstly, the gauge boson's energy and momentum are expressed in terms of the three 
relevant masses $m_2 = m(\charone,\,{\mgrey\neuttwo})$, $m_1 = m(\neutone)=m_{LSP}$, 
and $m_V = m_{W,\,{\mgrey Z}}$: \\[-2mm]
%%%%%%%%%%%%%%%%%%%%%%%%%%%%%%%%%%%
\begin{displaymath}
  E_V         = \frac{1}{m_2}\, \left(m_1^2-m_2^2-m_V^2\right)
  \hspace*{6mm} \mbox{and} \hspace*{8mm}
  |\vec{p}_V| = \frac{1}{m_2}\, \sqrt{\left(m_2^2-m_1^2-m_V^2\right)^2 - 4\,m_1^2\,m_V^2}
\end{displaymath}
%%%%%%%%%%%%%%%%%%%%%%%%%%%%%%%%%%%
From this the inverse expression has to be calculated for either the chargino 
{\mgrey(neutralino)} mass. Next, the effect of the finite boson mass width is 
added by convolusion with a Breit-Wigner function 
$\mbox{BW}\left(m_V^0, \gamma,\, m_V-\frac{\Delta m}{2},\, m_V+\frac{\Delta m}{2}\right)$ 
and, finally, some Gaussian smearing is added to the emerging fit function: \\[-5mm]
%%%%%%%%%%%%%%%%%%%%%%%%%%%%%%%%%%%
\begin{align*}
  \mbox{E-spec}^{\prime} & = \mbox{\sf\,bin}_{i} 
  \,             \cdot\,    \mbox{BW}\left(m_V^0,\, \gamma,\; m_V-\frac{\Delta m}{2},\; m_V+\frac{\Delta m}{2}\right) + \mbox{E-spec} \\
  \mbox{\sf\,bin}_i & = \underbrace{\frac{1}{2} - \frac{a}{2}\cdot \left(\frac{x\,m_2 \,-\, E_b\,E_V} {\sqrt{E_b^2 - m_2^2}\,\cdot\,p_V}\right)}
  \, \cdot  \underbrace{\left[\Theta_{{\left(\frac{R+1}{\sigma_{j}\,b}\right)}} - \Theta_{{\left(\frac{R-1}{\sigma_{j}\,b}\right)}}\right]
    \,+\, \frac{a\,b}{2\pi}  
    \left  (e^{-\frac{\left(R-1\right)^2}{2\,(\sigma_{j}\,b)^2}} -  e^{-\frac{\left(R+1\right)^2}{2\,(\sigma_{j}\,b)^2}}\right)}\,, \\%[-3mm]
  \,                & \hspace*{8mm}\mbox{no Gaussian smearing} \hspace*{30mm}\mbox{adds Gaussian smearing}
  % ---------------------------------------------------------------
\end{align*} \\[-9mm]
%%%%%%%%%%%%%%%%%%%%%%%%%%%%%%%%%%%
\begin{displaymath}
  \mbox{with:}\hspace*{5mm} 
  b = \frac{m_2}{\sqrt{E_b^2 - m_2^2}\,\cdot\,p_V}\,,\hspace*{8mm} 
  R = \frac{x\,m_2 \,-\, E_b\,E_V} {\sqrt{E_b^2 - m_2^2}\,\cdot\,p_V}\,,\hspace*{8mm} 
  \sigma_j \;\mbox{: jet energy resolution}\,.
\end{displaymath}
%%%%%%%%%%%%%%%%%%%%%%%%%%%%%%%%%%%

%%===========================================================================
%% page 4
%%========
\begin{figure}[!h]
  \setlength{\unitlength}{1.0cm}
  \begin{picture}(10.0, 5.65)
    % ---------------------
    \put( 0.00,-0.4)  {\epsfig{file=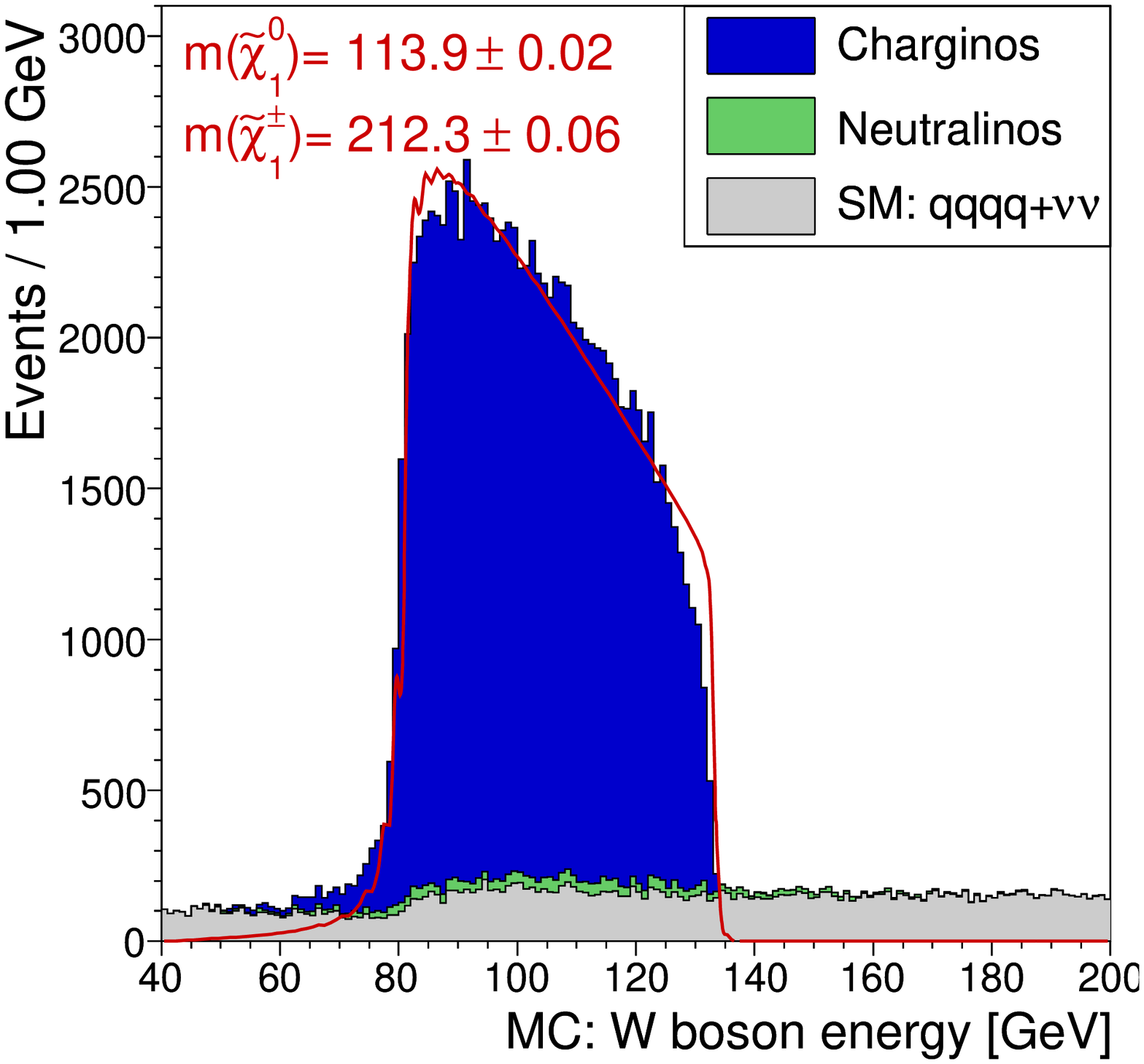, clip= , width=0.46\linewidth}}
    \put( 7.25,-0.4)  {\epsfig{file=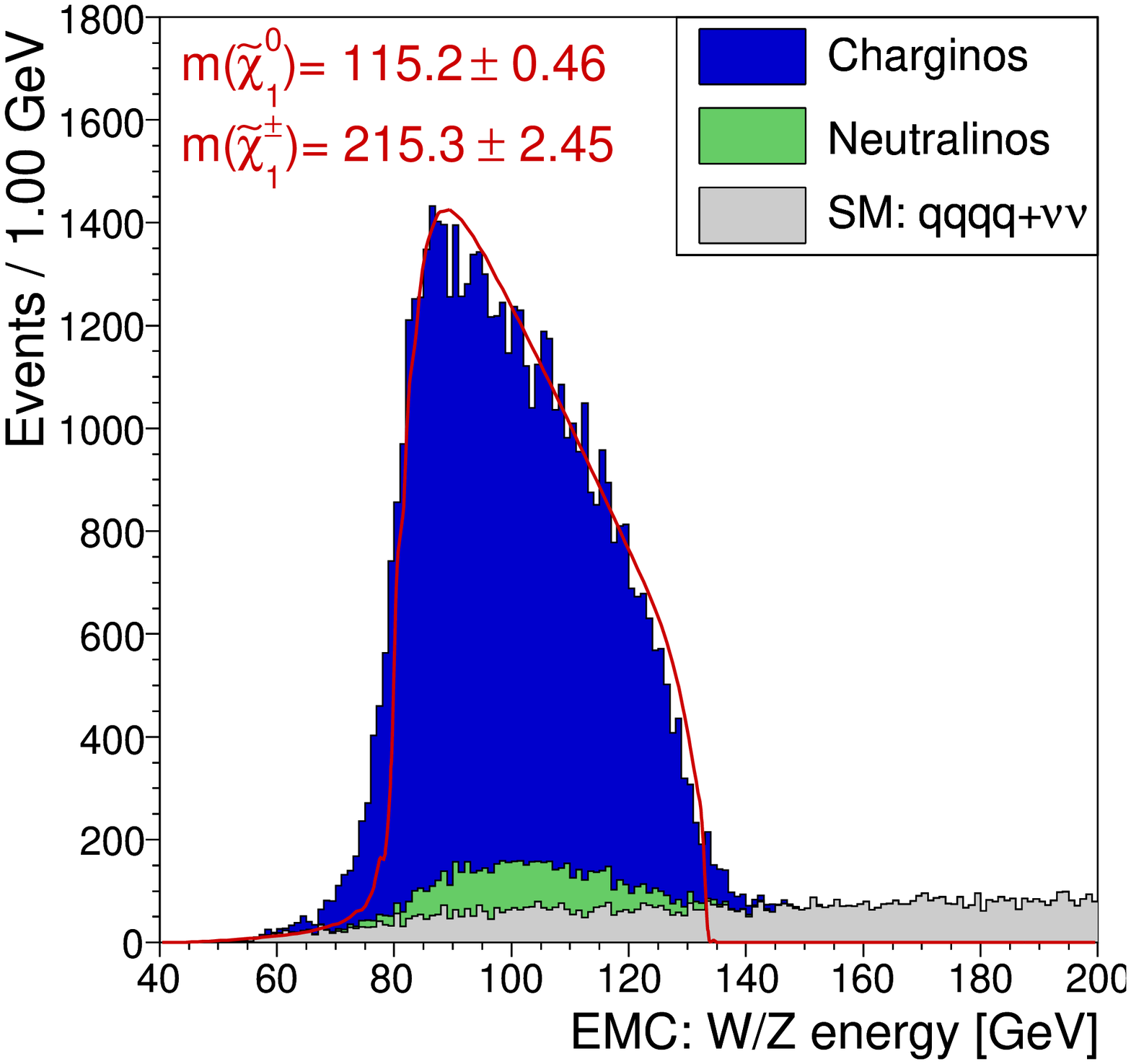, clip= , width=0.46\linewidth}}
    % ---------------------
    \put( 0.04,-0.1)  {(a)}
    \put( 7.10,-0.1)  {(b)}
    % ---------------------
  \end{picture}
  \caption{Fit to the energy spectrum of the $W$ boson (a) on generator level 
    and (b) after reconstruction and a first rough chargino selection.} 
  \label{fig:fitWene-dk}
\end{figure}
%------------------------------------
The fit values for the reconstructed bosino masses are actually quite close to 
the nominal generator level values: \\[-6mm]
\begin{align*}
  m(\charone) =\; & 215.3 \pm 2.45\;\,\mbox{GeV} \quad \lra \quad  216.5\;\,\mbox{GeV} \hspace*{6mm} \mbox{(nominal chargino mass)}       \\
  m(\neutone) =\; & 115.2 \pm 0.46\;\,\mbox{GeV} \quad \lra \quad  115.7\;\,\mbox{GeV} \hspace*{6mm} \mbox{(nominal neutralino/LSP mass)}.
\end{align*}

Although the error on the chargino mass is still large, this is expected to be reduced 
by a more sophisticated selection (higher $\varepsilon\times\pi$) and a simultaneous 
fit to the cross sections of both signal processes, i.e. chargino and neutralino 
pair production. The cross sections could then serve as a further input to the fit 
of the gauge boson energy spectra. 

While only the dominant SM background ($WWZ\to q\bar{q^{\prime}}q\bar{q}^{\prime}\nu\bar{\nu}$) 
is considered for the distributions in Fig.~\ref{fig:fitWene-dk}~(a,b), signal and 
background processes are normalised to the same luminosity of 500~fb$^{-1}$. This provides 
an estimate of the sensitivity of the chargino (neutralino) selection to a possible 
contamination from SM background.

\section{Conclusions and outlook}
Physics analyses using a full detector simulation play an important r\^ole in the 
ongoing optimisation effort for the ILC detector. This study, although performed on 
simulation data of a previous detector model (\texttt{LDCPrime\_02Sc}), constitutes 
one part of the general effort for the ILD detector concept. 
In cases where the gauge boson masses are reconstructed from 4-jet final states, 
the mass and energy resolution is expected to improve drastically due to the interplay 
between a highly segmented calorimeter with tracking ability for neutral particles 
and the new approach of Particle Flow algorithms. Together with more sophisticated 
fits to the resulting gauge boson energy spectra, mass resolutions of 0.5-1~GeV are 
achievable for those SUSY particles accessible to the ILC ($\charone$, $\neuttwo$, and $\neutone$).

Next steps will be to include all SM and relevant SUSY background, optimise the 
selection and fit procedures and utilise different polarisation scenarios to 
improve the S/B-ratio.

% ****************************************************************************
% BIBLIOGRAPHY AREA
% ****************************************************************************

\begin{footnotesize}
% ----------------------------------------------------------------------------
%
% IF YOU DO NOT USE BIBTEX, USE THE FOLLOWING SAMPLE SCHEME FOR THE REFERENCES

% ----------------------------------------------------------------------------
\end{footnotesize}

% ****************************************************************************
% END OF BIBLIOGRAPHY AREA
% ****************************************************************************

\end{document}